\documentclass[10pt,twocolumn]{article}
\usepackage{amsmath, amsfonts, amscd}

\begin{document}
\title{On topological defects}
\author{E.\ D.\ M.\ Kavoussanaki\thanks{e.kavoussanaki@ic.ac.uk} \\
Theoretical Physics, Blackett Laboratory, Imperial College,\\
Prince Consort Road, London, \\
SW7 2BZ, U.K.}
\date{\today}
\maketitle

{\sc{Abstract\ }}{\small{The language and methods of algebraic topology, particularly 
homotopy theory, have been extensively used in the study of the identification, the
classification and the evolution of defects. Topological methods provide the
means for the detection of gross features, such as holes in a manifold, and
therefore one would clearly anticipate that the defects, if identifiable by
such means, to be linked with that kind of configurations. In physical
media, however, defects usually manifest themselves in ways which have no
distinct and direct association with patterns of topological value along the
lines of our previous reference. In this paper we will postulate that the
defects do not themselves correspond to topological features. Instead, they
are forced to exist in order to redress the violation of physical rules
happening as a result of the topological nature of our system. We will, thus, assume that
the topology of our system changes with the introduction of defects. Therefore,
the relevent to our study topological methods will be applied to a well defined
physical system only in the absence of defects. 
}} \\
\\
An \textit{ordered medium} corresponds to some real smooth manifold $M$
where a function $f$ is defined which assigns to every point of the medium
an order parameter. Furthermore, over our medium, $f$ should be a continuous
function. For example, if $f:\mathbb{R}^{3}\rightarrow \mathbb{C}$ is a
continuous function throughout $\mathbb{R}^{3}$ then our medium is $%
\mathbb{R}^{3}$ and the order parameter is a complex number. In our
discussion, we will refer to that function $f$ by the name \textit{order
parameter}, even though this title might be somehow confusing\footnote{one, for
example, might consider the space of functions as that relevant to an
ordered medium and thus start all sorts of investigations on a different
sphere}.

Excluding the trivial case where the order parameter is constant throughout
the medium (which is called thus uniform), we will focus our attention to
non uniform media where the function, through connected space, varies
continuously apart perhaps (depending on the specific configuration) at
isolated regions \cite{mermin}.

Suppose $f$ to be an order parameter. Take $f:\tilde{M}\rightarrow \mathbb{C}
$ to be everywhere continuous apart from some region $M_d \subset \tilde{M}$%
. Obviously, according to our choice of $f$ we can force the existence of a
certain $M_d$ in $\tilde{M}$. We could consider the $M_d$ region as a hole
in $\tilde{M}$ so that $f$ is kept always continuous on a certain area of
its domain. The latter we will denote as $M$: $\tilde{M}-M_d$. In this way,
the order parameter will be assigning topological properties to our medium $M
$. However, this is an ambiguous technique in case $f$ is defined throughout 
$\tilde{M}$ since we will be just ignoring certain values of the order
parameter and thus be lead to a non equivalent situation. As we will see,
discontinuities in $f$ can be appropriately smoothed out so that any
measures of the kind outlined already will be unnecessary.

Primarily, our contemplation to consider order parameters with
discontinuities might seem quite far fetched to someone having in mind the
situation in physical systems where the order parameter is always
continuous. Its relevance will become evident if we introduced the notion of
a \textit{defect}. We will postulate that a defect corresponds to an
appropriate change in the order parameter, applied only to those regions
where originally it was discontinuous, having as a main aim to restore the
continuity of that function throughout the area where it was originally
defined. Thus, the discontinuities will indicate where the introduction of a
defect is needed. In this way our medium will become the same as the domain
of the order parameter diminishing the possibility of the previously
mentioned ambiguity.

As we will see, for a certain order parameter behavior, the defect induces
an \textit{invertible} change to that function. Thus, if we were to lift the
defect we would surely recover the underlying discontinuity. However, there
are situations for which the discontinuity happens artificially, and thus,
unless we remember the exact formula for the order parameter, any change
intended to make that function continuous can not be reversed. In physical
systems where the order parameter is everywhere continuous, defects are,
possibly, already there and therefore the continuity has been appropriately
restored. If, though, we were to remove those defects and recover the order
parameter configuration, we would surely see those discontinuities for which
this procedure is invertible.

It is, thus, inevitable to ask, how one identifies defects in continuous
configurations. Moreover, how one knows how the order parameter should
behave if one were to take them away. Here, we, thus, have to introduce the
notion of a \textit{preferred order parameter}. Since our medium is a
physical system, it should be described by certain equations the solution to
which is our order parameter. For various systems there exist many solutions
to the relevant equations (and here one should consider all possible
solutions including those having the order parameter constant). Each such
order parameter corresponds to a particular energy distribution over our
medium. We will assume that the solution preferred will be the least
energetic one. There could be situations where several order parameters will
be equally good in terms of their associated energy distribution. We will
postulate that our physical system does not distinguish between them.

Suppose that for a system in equilibrium (where we can safely identify
preferred order parameters) one finds that the corresponding order parameter
does not, over all, have a preferred configuration. If, in this particular
case one finds that there are some regions in our medium where the order
parameter deviates from the surrounding preferred configuration then one
should think of these areas as possible candidates for defects. If, further,
one lifts those configurations and seeks for a good replacement among those
functions having some physically explained priority then, in case such an
investigation turns out to be in vain, one can safely link these areas with
a defect. Here we have assumed that the way one discards certain - otherwise
preferred - order parameters for a given region has to do with whether they
violate the continuity requirement for the total order parameter defined
over all the manifold. Thus, order parameters which occur in contrast to the
natural tendency of the system, should underline the existence of defects.

To make the point clearer consider the situation where there are two
different order parameters which give the same energy distribution over all
the medium. Suppose that $f_{1}=c_1$ and $f_{2}=c_2$ with $c_1$, $c_2$
constants. Our medium is one dimensional. Take that the over all our
manifold order parameter is 
\[
f(x) = \left\{ 
\begin{array}{ll}
f_{1} & \mbox{if $x>0$} \\ 
c & \mbox{for $x=0$} \\ 
f_{2} & \mbox{if  $x<0$}
\end{array}
\right. 
\]
Obviously, at $x=0$ there will be a discontinuity of $f$ regardless of the
actual value of $c$. Thus, that discontinuity will be hidden if we
introduced a defect which would take us continuously from $c_1$ to $c_2$ or
vice versa. The order parameter configuration which we could choose should
satisfy the relevant equations for our system. This example, for those
familiar with our subject, monitors the situation of the one dimensional
kink. In that system, $c_1$ and $c_2$ correspond to the two minima of the
relevant potential. We took that, in the absence of a defect, those two
values for the order parameter would create a discontinuity at the $x=0$
point of our one dimensional manifold. Thus, introducing a defect there,
would smooth out our function in accordance to the relevant equations, and
form a continuous transition between those two minima.

Alternatively, if our relevant order parameter was 
\[
f(x)=\left \{ 
\begin{array}{ll}
f_1 & \mbox{for $x>0$} \\ 
c & \mbox{at $x=0$} \\ 
f_1 & \mbox{for $x<0$}
\end{array}
\right. 
\]
for some constant $c\ne f_1$ then that discontinuity at $x=0$, if removed by
forcing, for example, $f(0)=f_1$, would not be recoverable since there would
be no trace left to indicate what type of measure we took, and indeed if we
took any measure, in order to make the order parameter continuous. In that
particular situation we will assume that the domain of the order parameter
and $M$ are already the same even though there is a formula dependent
discontinuity at $x=0$. In such cases the $M_d$ could be considered to be
the empty set. Surely, there are other ways of making the order parameter be
continuous throughout $\mathbb{R}$. We will consider that the process of
smoothing-out should happen in a way that the least changes to $f$ would
take place.

Hereafter we will consider that if the order parameter, around any point $p$
of the manifold, has such a configuration so that the $\lim f$ at $p$ will
be well defined, then 
\[
f(p)=\lim_{p^{\prime}\rightarrow p} f(p^{\prime})
\]
regardless of the original value of the order parameter there. In the
converse case where the $\lim f$ at $p$ does not return a unique value, then
in order to restore the continuity - our primary aim - we need to smooth-out
our function in an appropriate way around the point $p$. Our assumption here
is that the order parameter will be exhibiting discontinuities only in the
absence of defects. Thus, $f$ is continuous \textit{because} of the
appearance of those configurations.

The idea of \textit{invertible} comes as a direct consequence of what has
been already stated. If for a given $f$ and a given $\tilde{M}$ we find that
on some $M_d$ there are discontinuities of that function being exhibited
then, our first step is to call $M:\tilde{M}-M_d$ our relevant domain. Then,
we will try and see if we can apply any preferred order parameter
configuration on $M_d$ so that $M\rightarrow \tilde{M}$ in the sense that
the new order parameter will be a continuous function throughout a medium ($M
$) equal its original domain of definition ($\tilde{M}$). If we are unable
to find any such preferred order parameter that will smooth-out our $f$
completely then the introduction of a new order parameter configuration will
be an invertible process. Otherwise, the additional alterations could not be
recovered.

Moreover, if we are given an order parameter configuration, take, for
instance, the former case described above (the kink), then we will be unable
to find if there are any defects there unless we are provided with the
information as to which are the order parameters that the system would
prefer to have.

Assume the planar spin case, also quoted by Mermin\cite{mermin}. We remind
the reader that the system of planar spins corresponds to a flat $2D$ medium
at each point of which a $2D$ vector is being assigned via the appropriate
order parameter. Mermin\cite{mermin} claims that if we removed the
information about the order parameter from inside a certain disk on that
plane, we would still be able to recover a discontinuity at the center of
that region in case the relevant vector winds by $2\pi n$ with ($n\ge 1$)
around it. That is obviously true \textbf{if} one is provided with the 
\textit{additional} information that our order parameter should be of a 
\textit{non zero magnitude} throughout that disk. If we were allowed to put
to zero the magnitude of the order parameter (thus, the preferred order
parameter could take the zero value) it is not necessary that we would be
able to recover \textit{any} discontinuity at the center. Thus, the
identification of areas where a defect is needed is the result of the
application of topological methods to a physical system which, we assumed,
tends to acquire only preferred order parameters.

It might seem, though, oxymoron the fact that the system does not take care
of the continuity of the order parameter itself, even though any
discontinuous behavior of that function cannot be physically allowed. It is
inevitable, thus, to require an answer to the question of how the system
manages to create a situation against its natural tendency. One, thus, needs
to clarify whether the order parameter takes a particular form because there
is some intrinsic value to the possibly created discontinuities or whether
this is just a trick to explain how the system chooses a state associated
with a certain continuous function $f$.


First of all, in order to create a discontinuity of the kind we already
outlined we need to have a system which has more than one preferred order
parameters. Furthermore, to simplify matters, we will assume that those
functions are bounded from above and below and that their range of values
correspond to entirely different sets. Suppose now that our medium $\tilde{M}
$ can be divided into independent sections denoted as $S_k$'s, for which we
have 
\[
\bigcup_{k=1}^{m} S_k \stackrel{\subset}{\_ }\tilde{M}
\]
and 
\[
B_{kk^{\prime}}=S_k\cap S_{k^{\prime}} 
\]
a part of the common boundary of $S_k$ and $S_{k^{\prime}}$ or, otherwise,
the empty set $\emptyset$. We will take that all the $S_k$'s around a
certain $S_{k_0}$ belong to the latter's environment and do not affect the
physics of that set. Thus, the $S_{k_0}$ can be thought of as an independent
physical system subject to no, by the environment imposed, boundary
conditions. If that happens for each $S_k$ of $\tilde{M}$ then the order
parameter will be allowed to evolve autonomically in each different section
of the manifold.

Even if our manifold $\tilde{M}$ is separated into physically independent
regions, we will assume that the parameterization in each of those areas
does not happen independently. Thus, there is a global chart $\psi$
assigning coordinates to each point of $\tilde{M}$. Further, the
identification of preferred order parameters will result from minimizing the
relevant potential. If the latter is given in terms of manifold coordinates,
then the order parameter configuration will be parameterization dependent
something which we wouldn't like to have. The reason is that we want to
allow the system to choose \textit{any} order parameter at each $S_k$ for
any given $\psi$. Thus, the potential will have to be given in terms of $f$,
the order parameter. It is essential to stress that by minimizing that
potential we will be finding order parameter values which can be applied in
different ways to our manifold. That is to say that the way the order
parameters we will be finding, should vary with manifold coordinates is not
determined and therefore the realization of the potential over some $S_k$
can have many different forms for a given $\psi$.

Suppose that 
\[
\bigcup_{k=1}^{m} S_k \mbox{\bf{=}} \tilde{M}
\]
and 
\[
B_{kk^{\prime}}=S_k\cap S_{k^{\prime}} 
\]
a part of the common boundary of $S_k$ and $S_{k^{\prime}}$. The boundary of
every $S_k$ is being thought of as a wall that prevents information about
the order parameter on its one side to cross to the other side. Assume that
the way to create such a situation is by externally forcing the system to
become fragmented.

Taking that the manifold is separated into independent regions each of which
corresponds to a physical system in its own right seems accommodating. Since
in each of those regions the order parameter is a continuous function, each
physical system will be related to a function with no discontinuities within
it. Thus, the whole $\tilde{M}$ cannot be considered as one physical system.
The latter will be implemented when the boundaries between the various $S_k$%
's fall and defects appear which smooth-out any discontinuities at the
connecting borderlines.

This approach lets us assume that those discontinuities are happening not
because of some intrinsic property of our physical system but instead
because of external intervention. They are, thus, forced to appear because
of the from the outside imposed fragmentation of our medium. This can be related
to what 
Polturak\ , Carmi\ \& Koren did in their recent experiment\cite{polturak} forcing a
predetermined domain structure. In this way the defects will be created
after that imposed structure is left to naturally evolve and smooth-out, in
an appropriate way, any initial discontinuities.

Another way we can create discontinuities is by assuming that the medium
consists of disconnected regions being at some distance $D$ away from each
other. Thus, 
\[
\bigcup_{k=1}^{m} S_k \subset\tilde{M}
\]
and 
\[
B_{kk^{\prime}}=S_k\cap S_{k^{\prime}}=\emptyset 
\]
At each $S_k$, being thought of as an independent physical system, the order
parameter takes a preferred configuration. One can postulate that those
regions either grow or come closer so that they become correlated. Thus, the
one feels the presence of the other, say after a distance $d$. Since then
the order parameter in each of those regions cannot acquire a form
independently. It could be possible that the form of the order parameter,
beyond that point, will not anymore correspond to a preferred one for either
of those regions. The defect, thus, will be appearing due to the interaction
between those areas.

This situation could be related to what has been performed already in
superfluid $^{3}He$. Using the fact that $^{3}He$ is an excellent neutron
absorber, two experiments, one in Helsinki\cite{helsinki} and another in
Grenoble\cite{grenoble} subject it to neutron bombardment to heat a small
region of the superfluid. Small bubbles of normal fluid appeared which, by
rapidly cooling down, evolved independently from the surrounding superfluid.
The order parameter of the surrounding superfluid $^{3}He$ could not follow
the changing temperature front fast enough\cite{bunkov}. Consequently,
internal regions of the hot volume transit into the superfluid phase
independently in accordance with the Kibble/Zurek cosmological scenario.

Here, thus, we have postulated that the main ingredient for the production
of defects is to create \textit{independent} regions within the manifold
where the order parameter acquires some preferred form. Moreover, we assumed
that in each one of those regions the phase transition takes place
individually. Hence, the phase transition by itself \textit{will not} create
those domains in our manifold.


E.D.M.K. would like to thank Dr. R. J. Rivers for fruitful discussions and
support.

\end{document}